# Wavelet Based Frequency Detection Using FPGAs

Caleb Hill and Darshika G. Perera

Department of Electrical and Computer Engineering, University of Colorado Colorado Springs

Colorado Springs, Colorado, USA

## Abstract

In the realm of signal processing, frequency and spectrum detection are fundamental tasks that can be computationally intensive. This project leverages the power of FPGAs to perform wavelet analysis on an input signal. The goal is to detect the presence of a specific frequency component - in this case, 6 kHz. Our experiments demonstrate that wavelet-based spectral detection is both possible, and easily implemented using an FPGA.

# SECTION 1 - INTRODUCTION

In the realm of signal processing, frequency and spectrum detection are fundamental tasks that can be computationally intensive. Traditional methods often involve the use of local oscillators and other complex hardware components. However, with the advent of Field Programmable Gate Arrays (FPGAs), it has become possible to perform these tasks more efficiently and flexibly. FPGAs, with their reconfigurable digital logic blocks, offer a powerful platform for implementing real-time signal processing algorithms.

This project leverages the power of FPGAs to perform wavelet analysis on an input signal. The goal is to detect the presence of a specific frequency component - in this case, 6 kHz. Wavelets, which are mathematical functions with a zero mean, allow for the decomposition of a signal into different frequency components. This process, known as wavelet transformation, produces a time-frequency representation of the signal, enabling the identification and study of specific frequency components over time.

The applicability and usefulness of predefined wavelets for spectrum decomposition of an unknown signal are demonstrated in this project. The ability to detect a specific tone in an audio signal, identify a particular frequency in a radio signal, or analyze the spectral content of any other type of signal, opens a myriad of applications. By studying the presence and behavior of the 6kHz component over time, valuable insights into the characteristics and behavior of the original analog signal can be gleaned. This project, therefore, stands at the intersection of hardware efficiency and signal processing, paving the way for innovative solutions in the field.

The motivation for this project stems from previous research in wavelet applications and the need to more rapidly implement them using real-time logic. Previously, simple wavelets have been applied to phase information of an incoming signal to detect transients indicating carrier frequency hops. These processes were extremely computationally intensive and represented the bulk of the compute time in the study [1]. Furthermore, expanding on study of the Analog to Digital Converter was a topic of prime interest.

The following report is divided as follow: Section 2 begins with an introduction and general coverage of wavelets, their design, and applicability. Next, the Analog to Digital Converter is described with particular emphasis on the capture of oscillating or time varying signals. The Section concludes with a description of the conceptual approach to designing this project which was taken before touching the hardware or writing any Verilog code. Section 3 covers the functionality and interworking of the various modules created to run on the hardware. Included in the module descriptions are issues encountered and their resolution. Section 4 concludes the report and shows the outputs of the system as designed to a real input signal.

SECTION 2 – PROJECT DESCRIPTION

Background: Wavelets

Wavelet Analysis and Transforms have gained prominence in signal analysis and are now ubiquitous in the areas of image and biomedical signal processing, geophysics, data compression, finance, and computer graphics, to name only a few [2][3][4]. Wavelets are mathematical functions that can be applied to separate data into different frequency components, and then study each component with a resolution matched to its scale. They have advantages over traditional Fourier methods in analyzing situations where a signal contains discontinuities and sharp spikes, and the preservation of temporal details is critical [5].

A wavelet is a wave-like oscillation with an amplitude that begins at zero, increases, and then decreases back to zero. It can typically be visualized as a "brief oscillation" like one might see recorded by a seismograph or heart monitor. Generally, wavelets are purposefully crafted to have specific properties that make them useful for task specific signal processing. Wavelets can be combined, using a "shift, multiply and sum" technique commonly known as convolution (eq 1), with portions of a unknown signal to extract information from the signal.

$$y[n] = \sum_{k=-\infty}^{+\infty} x[k] \cdot w[n-k] \qquad (eq\ 1)$$

In wavelet analysis, wavelets are convolved with the signal under analysis, allowing large-scale wavelets to reveal low-frequency harmonics, while small-scale wavelets reveal high-frequency harmonics. This makes them particularly useful for analyzing physical situations where the signal contains discontinuities and sharp spikes [3][5]. However, such analysis trades some advantages and disadvantages with traditional Fourier methods in time/frequency feature extraction of signals. Specifically, while Fourier analysis provides good frequency resolution but poor time resolution, wavelet analysis provides poor frequency resolution but good time resolution, particularly for high frequency analysis. In high frequency cases, multiple oscillations at the target frequency can be added to re-narrow the spectral bandwidth, but this comes naturally at the cost of time resolution or responsiveness.

Wavelet analysis uses reference basis functions, often called mother wavelets. Examples of different wavelets are shown in figure 1. The essential properties of the wavelets are their scale (Dilation) and location (Translation). The scale or dilation determines how stretching or compression of the wavelet with respect to a given axis. For one-dimensional signals, smaller scale values are shorter in time and thus capture high-frequency information, while larger scale values are longer in time and capture low-frequency information. Correlation of a signal and a wavelet basis function is revealed by their convolution. For shorter wavelets (scale or dilation is compressed), time resolution of the output convolution is the most precise, whereas longer wavelets (lower frequency) exhibit poorer time resolution[7][3][10].

There are several types of wavelets, commonly including the Morlet, the Mexican Hat, and the Haar. The Haar wavelet, named after Alfréd Haar who introduced it in 1909, is the simplest of all wavelets and is defined below in equation 2. A perhaps more recognizable wavelet is the Morlet Wavelet, also known as the Gabor wavelet, and is defined in equation 3. In the case of the Morlet wavelet, the characteristic symmetric growth and decay of the an oscillating signal is clearly identifiable. This particular wavelet is a versatile tool used in a variety of fields due to its excellent time-frequency resolution and multi-resolution analysis capabilities [2].

$$\psi(t) = \begin{cases} 1 & \text{if } 0 \leq t < 0.5, \\ -1 & \text{if } 0.5 \leq t < 1, \\ 0 & \text{otherwise.} \end{cases}$$

*(eq 2) - Haar Wavelet*

$$\text{Morelet} = e^{i \cdot 2\pi \cdot \text{freq} \cdot t} \cdot e^{-\frac{t^2}{2 \cdot \left(\frac{\text{width}}{\text{freq}}\right)^2}}$$

*(eq 3) - Morlet Wavelet*

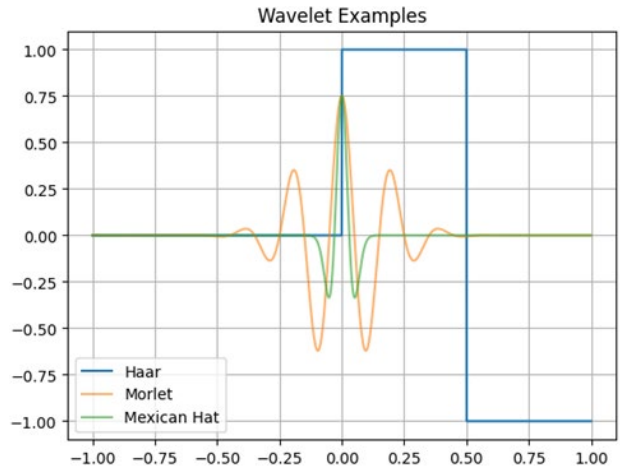

*Figure 1 - Wavelet Examples*

In medicine, it's used in magnetic resonance spectroscopy imaging where it offers an intuitive bridge between frequency and time information, clarifying the interpretation of complex head trauma spectra. It's also used to discriminate abnormal heartbeat behavior in electrocardiograms (ECGs) since the variation of the abnormal heartbeat is a non-stationary signal, making it suitable for wavelet-based analysis [6]. In geophysics, the Morlet wavelet has been used for numerous studies including tropical convection, the El Niño–Southern Oscillation (ENSO), atmospheric cold fronts, central England temperature, the dispersion of ocean waves, wave growth and breaking, and coherent structures in turbulent flows [7]. In signal, image, and video processing, it's applied in areas such as image compression, detection of microseismic signal arrivals, acoustic signal compressions, and solving integral and differential equations [8]. In neuroelectrics, Morlet wavelets are frequently used for time-frequency analysis of non-stationary time series data, such as neuroelectrical signals recorded from the brain [6]. The width of the Gaussian that tapers the sine wave is a crucial parameter of Morlet wavelets [9].

In this application, the Morlet wavelet will be exploited to analyze an unknown signal captured by an analog-to-digital converter (ADC). The ADC will convert an input analog signal into a quantized digital signal that can be processed by the Xilinx Spartan 3e FPGA. The Morlet wavelet, with its excellent time-frequency localization properties, will be used to decompose this input digital signal into different frequency components. By adjusting the parameters of the Morlet wavelet, focus will be on the spectral content near 6 kHz. The wavelet parameters and definition in this case are described by equation 3 and shown in figure 2. Continuous convolution of the discrete input signal will produce a sample time accurate response to the presence of spectral content at the target frequency. It should be noted that the wavelet definition describes a complex signal; that is, one with real and imaginary parts. This complex nature allows the convolution with a real signal to produce outputs similar to that from the output stage of a Software Defined Radio, which has been the target of previous studies [10].

Software Defined Radio (SDR) represents a significant shift in the evolution of radio technology, particularly from the standpoint of broadband reception. Unlike traditional hardware-based radio receivers which demodulate at a particular carrier frequency to reproduce the baseband signal (e.g., the voice signal sent over the radio), SDRs sample the incoming spectrum at rapid, periodic intervals and produce an output bit-stream which is passed to software for processing and demodulation. The sampling is accomplished using a Quadrature Receiver, which takes two samples of the instantaneous value of the signal that are 90-degrees out of phase. The 90-degree separation is accomplished by multiplying the instantaneous input signal by Cosine and Sine functions and sampling the resulting values to produce what are referred to as the In-Phase and Quadrature samples, otherwise known as "I/Q Samples". In accordance with the concept of Nyquist limits as discussed later, the sampling rate of the SDR determines the maximum bandwidth that can be analyzed, permitting one-half the sampling rate above and below a given center frequency. Positive and negative spectral content is possible because of the Quadrature, or complex receiver. Where this differs from the wavelet approach explored here is the SDR actively generates the cosine and sine signals to produce the I and Q channels. The wavelet approach by contrast permits I and Q components to be pre-generated, and the incoming signal is convolved with each, one sample at a time. This is essentially analogous to quadrature mixing in the SDR but does not require an active oscillator.

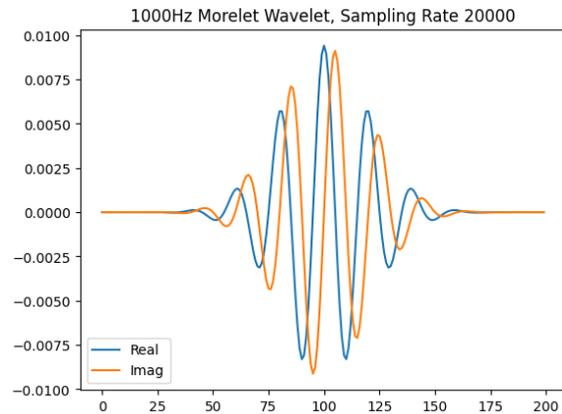

*Figure 2 - Example Complex Wavelet*

Background: Analog to Digital Converter

The Spartan 3E FPGA features an Analog-to-Digital Converter (ADC) and a programmable gain pre-amplifier which can be used in conjunction to capture an analog input signal for processing. The ADC is a two-input, Serial Peripheral Interface (SPI) based component that can play a crucial role in converting real-world analog signals into digital data that the FPGA can process. Each output is 14-bits per sample, permitting a possible 16,384 different values to represent a sampled input. The samples are signed integer values, and therefor correspond to the range of -8192 to 8191 over the dynamic range of the ADC [11][12].

Integral to the ADC, the programmable gain pre-amplifier is a vital component that amplifies the signal before it is digitized by the ADC. This amplification adjusts the input signal to maximize the quantization of the ADC over its dynamic range, particularly for low-voltage inputs, and can be adjusted based on the requirements of the specific application. Together, these components enhance the Spartan 3E FPGA board's capability to interface with and process a wide range of signals, making it a powerful tool for digital system design and signal processing. In this study, the input signals are adjusted to suit the minimum amplification required by the pre-amp.

The ADC on the Spartan 3E operates within a specific voltage range. It uses a reference voltage of 1.65 volts to determine the digital representation of the analog input signals. The ADC can accept an input voltage range from 0.4 volts to 2.9 volts. Values below 1.65 are considered negative, whereas values above 1.65 are considered positive. This range is the "window" within which the ADC can accurately convert an analog signal to a digital one. Any voltage value outside this range will be "clipped" to the maximum or minimum quantizable value and thus not accurately represented. The maximum input voltage that the ADC can handle without potential damage is 3.3 volts [12].

The gain setting of the pre-amplifier becomes crucial when dealing with inputs that don't deviate very far from the reference voltage. For example, if an input only varies by 0.025 volts, the 14-bit quantization can only represent the maximum value with a resolution of 163 steps. This is where the programmable gain pre-amplifier can help. By increasing the gain setting, the pre-amplifier raises the low voltage input signal to bring it closer to the ADC's maximum input range, thereby reducing quantization error. The example of 0.025 volts could be multiplied by a gain of 100 (negative, as implemented) to increase the quantization resolution to 16384 steps. It is important to note that the preamplifier is an inverting circuit as well. Negative signals - those below the reference voltage – are represented as positive integers, whereas positive signals are represented as negative integers. At a gain settings of -1, the minimum input of 0.4 volts or lower will yield 8191 as output from the ADC. Similarly, the maximum input at or above 2.9 volts will yield -8192. These are important characteristics to note when testing the functionality and implementation of the ADC [11].

The maximum sample rate of the Analog to Digital Converter is approximately 1.5 Megasamples per second (Msps). This is determined by the on-board clock, the sample data payload, and the nature of the Serial Peripheral Interface which transfers the data. The ADC takes a sample when instructed by assertion of the ADCON line, and both channels of the ADC simultaneously collect and buffer the input signal, totaling 28-bits. On the following assertion of ADCON, the signal is evicted from the buffer via the SPI one bit at a time. The signal is padded by two bits on the leading and trailing edges of the bitstream, and two bits between the channel A and channel B data, totaling 34 bits that must be offloaded. Taking the maximum clock rate into consideration and noting that 34 cycles will be required to serially transfer all bits, the maximum sample rate can be shown to be 1.47 Msps. This is an important limit to observe in experimentation and prototyping so as not to exceed the Nyquist limit for input frequency.

The Nyquist limit, named after Harry Nyquist, is a fundamental principle in the field of digital signal processing and sampling theory. It states that to accurately represent a continuous signal, the sampling rate must be at least twice the highest frequency component present in the signal. This is because a minimum of two samples per period are required to accurately capture the signal's oscillating characteristics. In this case, given a maximum sampling rate is 1.47 Msps, the Nyquist limit is approximately 0.735 MHz. Any frequency components above this limit will not be accurately represented and will lead to a phenomenon known as aliasing. Aliasing occurs when higher frequencies are incorrectly interpreted as lower frequencies. A common example of aliasing is the appearance of slowly moving or backwards spinning rotor blades in videos of helicopters due to the true rotational speed exceeding the video frame rate. Therefore, when sampling a signal, it's crucial to ensure that the sampling rate is at least twice the highest frequency you wish to capture. This adherence to the Nyquist limit allows for accurate signal representation and helps avoid the undesirable effects of aliasing.

As noted above, the ADC inputs in this study are controlled and set such that the gain of the pre-amp can be minimized to -1. Specifically, a sinusoidal signal with a DC bias of 1.65 volts will be used as the input. The amplitude of the sinusoidal component will be set to 1.25 volts in order to maximize the ADC quantization. This approach better enables the evaluation of the FPGA ADC in wavelet-based frequency decomposition analysis of input signals.

## Conceptual Approach

The initial design and theoretical testing for this examination was performed using signal analysis tools in python. This was conducted in three stages. First, the concepts of spectral identification were explored and refined entirely in simulation using native data types of 32-bit floating point providing extremely high resolution and minimal quantization error. Second, wavelet coefficients previously generated were converted to 14-bit integer values to enable compatibility with the Spartan 3e ADC output. Third, final testing was performed using entirely 14-bit simulations to validate functionality with reduced signal resolution.

Conceptual testing was based on close approximation of a high-resolution model to theoretical response of analog waveforms. Specifically, a wavelet function was coded based on the specifications of equation 3 (above). This wavelet approach was intended to function similarly to that of an SDR as described above, with the key difference being the arrangement of the complex multiplication. Where and SDR would output in-phase and quadrature signals that would each be convolved with a real-valued wavelet, this approach takes a real-valued sample stream and convolves it with the in-phase and quadrature coefficients of a pre-generated complex wavelet. This was specifically verified ahead of time using python to ensure the wavelet generation was functioning according to theory. To do so, a complex wavelet was generated and then convolved with real-signals at static frequencies to verify that only those with frequencies closely aligned to that of the wavelet would produce significant response. For additional analysis, "chirp" signals were generated and passed through the wavelet to show a signal with varying frequency in time will produce outputs only when spectral content is present that the wavelet is naturally responsive to. An example of the chip signal response is shown in figure 3.

Because the input to the wavelet was intended to be sequences of 14-bit values, the wavelet needed to be scaled accordingly. Accomplishing this scaling is relatively straightforward. All 32-bit real and imaginary coefficients (in-phase and quadrature components) of the wavelet that were generated in python were simply multiplied by 8191 and taken as integers. This simply trimmed any remaining fractional value of the coefficients. Rounding to the nearest integer was not considered necessary as any gain in detail would be less than the 14-bit resolution already provided.

Finally, testing of the concepts was performed again using entirely 14-bit simulation data. Specifically, all test signals were quantized to 14-bit values using the same method described above, and 14-bit the wavelet coefficients were applied. These results (shown in 5) indicated that the FPGA performing wavelet convolution with an incoming 14-bit signal should produce similar responsiveness as the higher resolution simulations. Thus, it was shown this concept should work using the on-board ADC provided the FPGA resources were sufficient to handle the wavelet coefficients and their convolution. Further, python simulations using 14-bit converted data permitted estimating real level of the wavelet response, and thus enabled setting the threshold values before even touching the hardware. The thresholds are noted by the gray horizontal lines of figure 5.

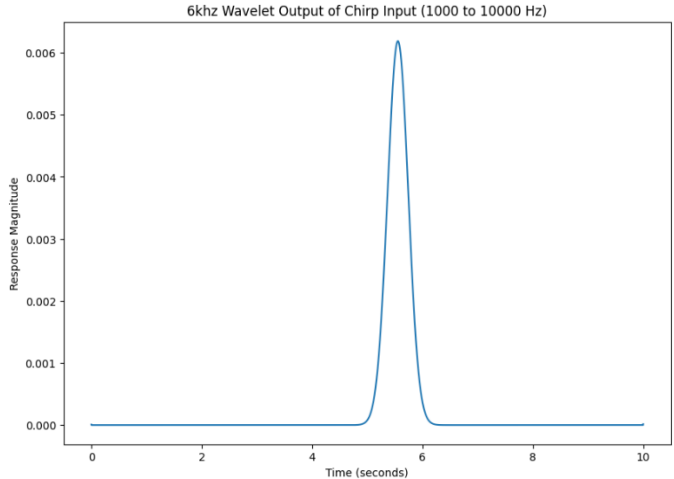

Figure 3 - Wavelet Response to Chip Signal

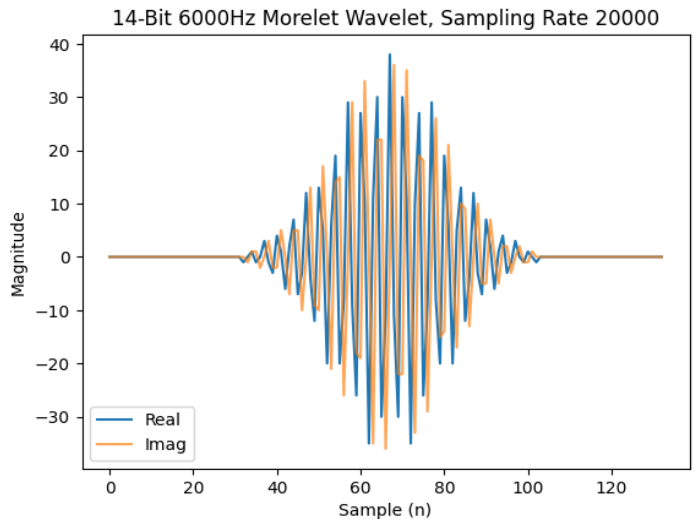

Figure 4 - Wavelet Applied in This Project

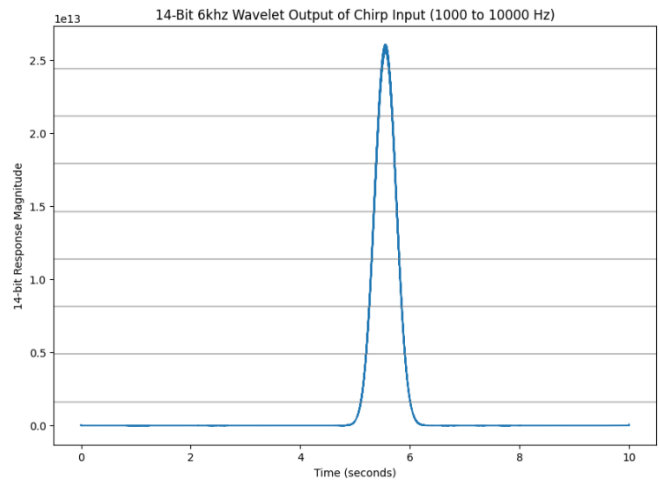

Figure 5 - Simulated Response of 14-Bit Chirp

SECTION 3 – HARDWARE DESIGN AND INTERWORKING

The hardware design approach revolved around two primary modules and three supporting modules. The primary components consisted of a modified Analog to Digital Converter module as initially designed in homework assignment 7, and a "wavelet filter" module designed to receive samples from the ADC module, place them into shift-register, convolve them with the wavelet coefficients, and calculate the magnitude-squared response of the convolution with each new sample. The supporting modules included a clock divider (Xilinx IP), a counter-based pulse generator, a delayed-pulse generator, an LED control module, and a debouncer module. The supporting modules are described first as they provide some context to the operation of the primary module structure.

Clock Divider: The intended sampling rate for the ADC was set to 20 Ksps to provide sufficient smoothing for a wavelet target with an initial target frequency of 1 KHz. This rate was maintained through the design despite the target frequency being shifted up to 6Khz in order to reduce the number of coefficients that had to be loaded by the FPGA and avoid potential resource limitations preventing completion of the design. The input 50 MHz clock was divided by five using the Digital Clock Manager (DCM) to produce a 10 MHz reference for the design. This provided ample speed, but also reduce the counter requirements to produce a 20 Ksps rate from the ADC. The DCM is the only Xilinx IP used in this study.

Counter-Based Pulse Generator: The 20 Ksps sampling rate was achieved by asserting the ADCON signal high at the desired rate and allowing the previously designed ADC master to simply free-run at the same speed. The module was created with a with inputs "clk", "reset", and "do run" and a single output "pulse". The design was based on a two-state finite state machine. Upon reset, the FSM defaults in the IDLE state and awaits a "do run" signal. When the "do run" signal is received, the FSM advances to a RUN state and begins counter-based pulse generation. A counter was created and synchronized to the input clk signal which was assumed to be the 10 MHz DCM output described above. The counter iterates 500 times before generating a one-cycle pulse and automatically returning to the start value. The result was a pulse every 50 µS, or at a frequency of 20 KHz. This output pulse signal is used to drive the ADC at the desired rate.

Delayed-Pulse Generator: The gain setting of the pre-amplifier was statically set to -1. Thus, rather than requiring multiple button presses to program the pre-amplifier before operation, it was desired to automatically set the gain upon system reset (via reset button). To accomplish this, a delayed signal is generated to provoke the ADC module to send the programming instruction. Specifically, the delayed-pulse generator module is built with inputs "clk" and "reset", and a single output called "pulse". The clk input is passed the 10 MHz DCM output described above. The module contained a single "always" block that loop three times at 100 Ns per loop, then assert a pulse signal for 1 cycle and simultaneously raise a flag that the pulse has been generated before ending the pulse signal. Thus, after 300 Ns, a 100 Ns pulse is generated which is used to indicate that the ADC module should program the pre-amplifier to the pre-set gain.

LED Control Module: In order to visualize the output of the wavelet module to the input signal captured by the ADC, the LEDs are used to progressively turn on as the response magnitude-squared crosses specific thresholds. The threshold values were determined in the experimental stage with 14-bit simulation data measuring the maximum response of a full-range signal at the target frequency. The maximum value was divided by 8 to segment the dynamic range into 8 divisions for the 8 LEDs onboard the Spartan 3e, and the threshold values were shifted down by a further one-sixteenth the max response value so place the final thresholds in the middle of each range.

The led control module was designed to receive the wavelet convolution output as a 50-bit signed number and compare it against the thresholds in a series of IF statements. The thresholds were likewise stored as 50-bit signed values. The initial design had this module wrapped by the Top Level module and data was passed from the Wavelet module to the LED module. However, it was observed that the comparison operations in the IF statements were not functioning properly, and 50-bit signed values passed from the Wavelet output to the LED module were being interpreted as max value 50-bit signed numbers for any value greater than 0 assigned. If the output was simulated and assigned form inside the LED module, it maintained value correctly. Therefor an initial change was to pull the LED module code inside the Wavelet module to avoid the requirement to exchange large numbers between modules. This did not completely resolve the issue however, and after considerable testing using ChipScope, it was decided to test smaller values by clipping the 32 least-significant bits (LSBs) from the Wavelet response output and from the threshold values. Clipping the LSBs resolved the issue. This was actually the final modification to the system and resulted in a fully working project model. Therefore, the LED control module code was left inside the Wavelet module to preserve the functionality of the system.

Debouncer: When a button is pressed or released, it doesn't make a clean, single transition from one state to another. Instead, the mechanical nature of the switch causes it to oscillate rapidly between states before settling, leading to multiple transitions. This can be interpreted as multiple button presses by the digital logic circuit, leading to erroneous behavior, especially if operating at relatively high clock rates. A debounce circuit mitigates this issue by ensuring that only a single transition is registered during each press or release, thereby providing a clean, bounce-free input to the logic circuit. This ensures accurate and reliable operation of the system. The deboucer used here was the same module used in all previous homework assignments and is based on dual flip-flop topology. Note, one-shot circuits were not needed in this design and thus not implemented in the final or functional code.

Pre-Amp and ADC Module: Interfacing with the pre-amplifier and analog to digital converter is accomplished using a modified code set from homework assignment 7. The code is largely unchanged, with a few small but critical modifications. First, despite indications from the iSim simulator that waveform generation to communicate with the pre-amp and ADC were correctly implemented, ChipScope analysis showed that the SPI clock was not properly returning to zero following programming instructions. This appears to have not caused an issue when measuring single DC values, as was done in assignment 7. However repeated sampling was not returning valid signals. Therefore, some state machine timing was adjusted and a flag was added to ensure the SPI Clock returned to zero following the pre-amp programming instructions.

The second issue concerned reading the ADC response. The initial implementation utilized a read-counter (decrementing) and iteratively assigned a register bit-values by array position as determined by the counter. The counter was synchronized to the input clock and decremented on the positive clock edge if the SPI Clock and an "enable get count" flag were both high. This worked without issue when the input clock was divided by a factor of 16. (the SPI clock is driven at half the input clock rate.) However, it is believed that by increasing the input clock rate from about 3.125 MHz to 10 MHz, there was not always sufficient time on the positive clock edge where the SPI clock was still positive, and thus bits count iterations were being lost and the ADC data corrupted. The read function was therefor changed to function as a shift register and assign a new LSB from the SPI MISO line on the negative SPI Clock edge. Because the SPI clock was not free-run, maintaining count was not necessary. This simple – yet critical – change resulted in correct data capture off the ADC.

As described above briefly, the pre-amp and ADCON (get sample) functions were no longer controlled directly by human interface and instead automatically set based on the system state. The interface lines remain unchanged in the pre-amp and ADC module, but their control is handled differently. The "do send command" as used in homework assignment 7 is now automatically asserted via the delayed-pulse generator as detailed above. The gain value command is statically set in the module code to 0x01 for a value of -1 since the oscillator amplitude output is controlled to fit within the ADC dynamic and quantization range. The "start read" command, also as used in homework assignment 7, is "flashed' automatically via the counter-based pulse generator at a rate of 20KHz to produce an output bit stream of 20 Ksps that is passed to the Wavelet (and LED) module.

Function testing of the modified ADC module was performed by providing a DC value with a known output. In this case, the maximum and minimum inputs were provided to produce 1FFF and 2000 outputs, respectively. The pulse generator was then free run to cause the ADC to repeatedly sample the DC values and the results were examined using ChipScope, as shown in figures 6 and 7.

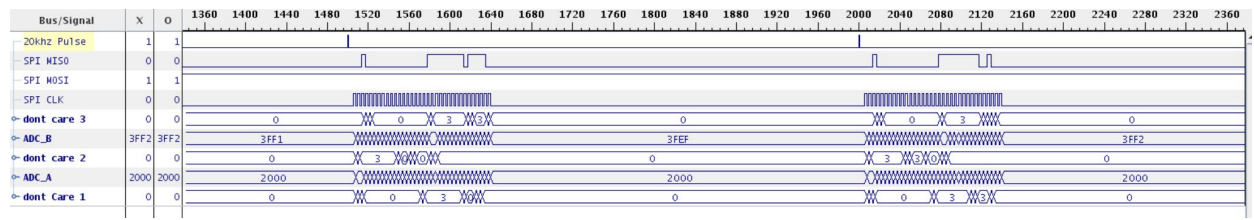

*Figure 6 - ADC Result of 2.9 Volt DC Input*

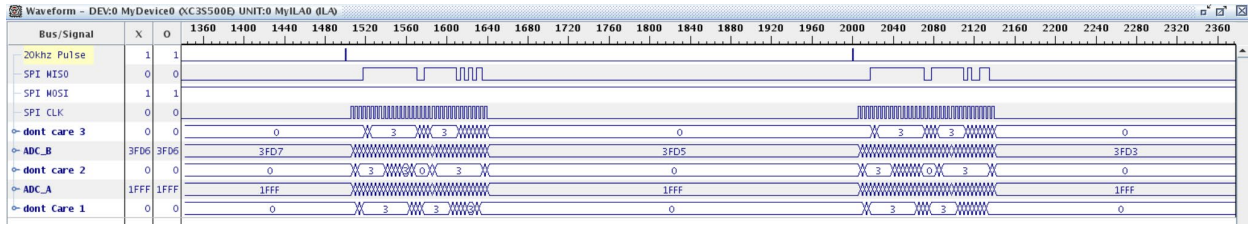

*Figure 7 - ADC Result of 0.4 Volt DC Input*

Wavelet Module: The primary functionality of this project is accomplished in by the application of the 14-bit stream of samples to the wavelet designed in the conceptual approach phase using python. The wavelet module currently also hosts the LED controller, but this is likely easy to slightly restructure so that the LED control functions are a separate module. However, as the code is in working form now, no further changes were made.

The Wavelet module concept of operation is to define the coefficients for the wavelet as defined by the 14-bit conversion in python and treat it like a Finite Impulse Response (FIR) filter. No Xilinx IP was used to this end, as the desire was to make the code "portable" to other FPGAs in the future (such as the Spartan 7). Additionally, there was strong desire to learn the detailed workings of building an FIR design. The coefficients are stored in Verilog Header (.vh) file that was generated by custom python script. This permitted much faster code completion and easy re-definition of coefficient values if needed. The number of filter coefficients and taps in the implemented design is 133. Thus, both the coefficients and the taps constitute a length 133 array of 14-bit register. At the sampling rate of the ADC (as dictated by the counter-controlled pulse generator), new samples are received from the ADC and placed into the least-significant slot of the taps array and all other taps values are shifted left. The entire taps array is piecewise multiplied by the entire coefficients array, and the products are summed for the result. This is performed for both the real and the imaginary coefficients and stored into a "real part" and "imag part" register.

The registers for the real and imaginary parts were both declared as 33-bit signed values to guarantee no overflow errors in the case of maximum response. Calculating the magnitude-squared response from these two parts simply involves squaring each, and then summing. Magnitude-squared was opted over the standard magnitude function as given by the Pythagorean Theorem because the inclusion of a square-root function added complexity unnecessarily. The actual value of the response isn't important so long as strong correlation of the input signal spectrum and the wavelet produces a large response, and weakly correlated inputs produce little response. However, maintaining and comparing the squared values required large registers to properly represent the numbers. 50-bits are allocated to the wavelet response. This large value was shown to cause some problems.

The wavelet module originally passed the full 50-bits to the LED module output-to-input interfaces handled by the top-level module. The LED module compared the 50-bit input against eight 50-bit thresholds that were defined as local parameters. As detailed above, the threshold values were determined in the experimental stage with 14-bit simulation data measuring the maximum response of a signal at the target frequency. The maximum value was divided by eight to segment the dynamic range into eight divisions for the each of LEDs onboard the Spartan 3e, and the threshold values were shifted down by a further one-sixteenth the max response value so place the final thresholds in the middle of each range. If the 50-bit input was above a given LED's threshold, a flag for that LED would turn be set to HIGH. The flags for each LED are assigned to the wire-outputs corresponding to the LEDs in the top-level module. The intent is to progressively turn on more LEDs as the response grows larger, and thus visually demonstrate the wavelet response to the designed target frequency. However, direct comparison of the 50-bit values always yielded a TRUE output, and turned the LEDs on, regardless of the actual response.

Through experimentation it was observed that any value greater than zero assigned to the response register was returning a TRUE if evaluated for being greater than any threshold. Verifying with ChipScope, all registers were properly treated as signed values, no overflow was present, and statically assigning a response signed value of the same bit-length was shown to function correctly. In troubleshooting, the LED controller code was moved inside of the Wavelet module to remove the top-level interfacing from the process. This did not make any difference. Further experimentation and troubleshooting led to an attempt at clipping the least significant bits from both the 50-bit response and threshold values. The 34 least significant bits were removed from the thresholds (which were all zeros by definition) resulting in a 16-bit values. The response is still defined as a 50-bit value, but the comparison operation only takes bits 49~32, leaving only an 18-bit number for the response. (The added bits are to ensure no overflow in the dynamic response). This solution was shown to work immediately, and the entire system and project was functional as desired.

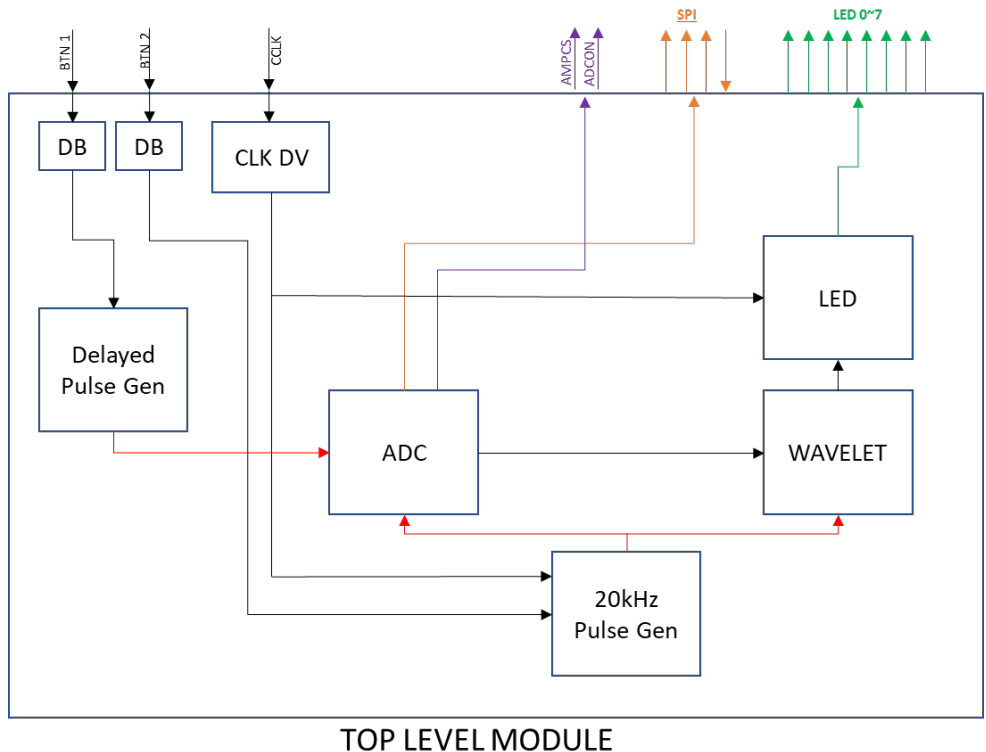

*Figure 8 - Module Architecture*

SECTION 4 – DISCUSSION AND CONCLUSION

This experiment demonstrates that wavelet-based spectral detection is both possible, and easily implemented using an FPGA. Utilizing the design described above, the LED response vs input frequency is shown in figure 9, and clearly demonstrates the wavelet-based response to the spectral content of the input. In this case, the input is a sinusoid swept across the responsive range of the wavelet.

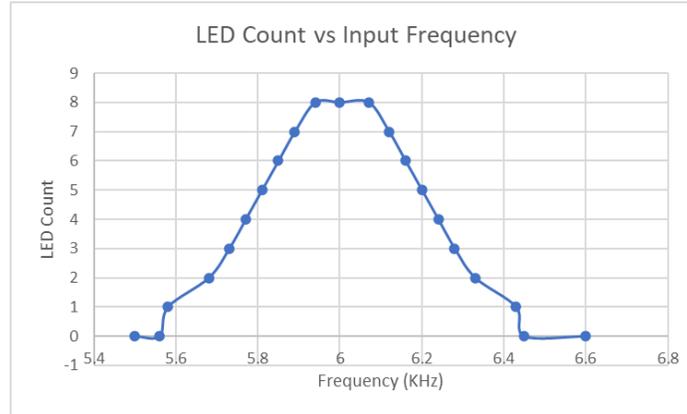

*Figure 9 - LED Activation vs Input Frequency*

Beginning the design with simulations based on Python signal processing techniques was shown to significantly speed up the overall project progression and making the project flow from Concept → Theoretical and Conceptual Testing → FPGA Prototyping truly a "rapid Prototyping" process. However more complicated and thus more real-world applicable designs are likely limited using the Spartan 3e due to limited resources available on the aging 3-series.

As can be seen from figure 8 below, the resource utilization is rather significant in this experiment. This was a known limitation on project initiation, and the ultimate reason for option to detect the single-frequency response and progressively display the result instead of early concepts oriented at using all LEDs to detect different spectral content. The 6-Khz wavelet-based detector required 133 real and imaginary coefficients, totaling 266. As detailed in the description of wavelets, lower frequency detection requires longer wavelets, which equates to more coefficients and more taps. Reducing the single wavelet used in this experiment to 1-Khz would increase the total coefficients to 1596, which would vastly exceed the available resources. Therefore, utilizing this hardware for more complicated, multi-frequency detectors may require reducing the bit depth of the ADC samples, shortening the wavelets and thereby increasing their frequency response range, or using multiple FPGAs to cover a wider spectrum.

| Device Utilization Summary (estimated values) | | | | [-] |
|---|---|---|---|---|
| Logic Utilization | Used | Available | Utilization | |
| Number of Slices | 3966 | 4656 | | 85% |
| Number of Slice Flip Flops | 1421 | 9312 | | 15% |
| Number of 4 input LUTs | 7035 | 9312 | | 75% |
| Number of bonded IOBs | 20 | 232 | | 8% |
| Number of BRAMs | 3 | 20 | | 15% |
| Number of MULT18X18SIOs | 20 | 20 | | 100% |
| Number of GCLKs | 5 | 24 | | 20% |
| Number of DCMs | 1 | 4 | | 25% |

*Figure 10 - FPGA Resource Summary*

The issues in comparing large bit value registers are still perplexing. Some potential reasons found in preliminary searches included resource limitations, in that the FPGAs have finite logic elements (LEs) and such large comparisons might exceed the available LEs. Timing constraints could also factor in, as the large bit-width registers require more logic levels to execute the comparison. Coding issues were mostly ruled out by testing using a variety of web-based Verilog compilers, and the issues being bypassed by simply reducing the bit-width of the registers in the comparisons. However, all three will warrant future research.

Additional future work with these wavelet concepts will likely include simple frequency shift keying demodulators. In fact, if bit depth from the ADC were reduced and wavelets shortened to the minimum length necessary to divide the sample space spectrum into two parts, a simple binary frequency shift keying demodulator may be possible on the Spartan 3e. However future work will likely be conducted with the Spartan 7 series boards and newer software sets like Vivado.

The opportunity to select, conceptualize, develop and prototype this project was fantastic capstone to the course. This project utilized concepts from every lecture, and every assignment, and leveraged code created in assignment 7 to produce an entirely new functionality. And, the entire process was done "Rapidly".

This work is inspired by the digital design research group at UCCS. This group has done extensive work in FPGA-based architectures, techniques, and associated models. Their analyses [13],[14] shows that FPGA-based systems are currently the best option to support complex applications and algorithms, such as the ones presented in this report, i.e., Wavelet Based Frequency Detection. Also, their previous work on FPGA-based accelerators, architectures, and techniques for various compute and data-intensive applications, including data analytics/mining [15],[16],[17],[18],[19],[20],[21],[22],[23],[24]; control systems [25],[26],[27],[28],[29],[30]; cybersecurity [31],[32],[33]; machine learning [34],[35],[36],[37],[38],[59]; communications [39]; edge computing [40],[41]; bioinformatics [60]; and neuromorphic computing [42],[43]; demonstrated that FPGA-based systems are the best avenue to support and accelerate complex algorithms.

Also as future work, we are planning to investigate hardware optimization techniques, such as parallel processing architectures (similar to [44],[45],[46],[36]), partial and dynamic reconfiguration traits (as stated in [47],[48],[49]) and architectures (similar to [50],[51],[52],[32]), HDL code optimization techniques (as stated in [53],[54]), and multi-ported memory architectures (similar to [55],[56],[57],[58]), to further enhance the performance metrics of FPGA-based Wavelet Based Frequency Detection, while considering the associated tradeoffs.